# Closing the gap between software engineering education and industrial needs

Vahid Garousi, Wageningen University
Görkem Giray, Independent Researcher
Eray Tüzün, Bilkent University
Cagatay Catal, Wageningen University
Michael Felderer, Blekinge Institute of Technology, and University of Innsbruck

**Abstract:**

According to different reports, many recent software engineering graduates often face difficulties when beginning their professional careers, due to misalignment of the skills learnt in their university education with what is needed in industry. To address that need, many studies have been conducted to align software engineering education with industry needs. To synthesize that body of knowledge, we present in this paper a systematic literature review (SLR) which summarizes the findings of 33 studies in this area. By doing a meta-analysis of all those studies and using data from 12 countries and over 4,000 data points, this study will enable educators and hiring managers to adapt their education / hiring efforts to best prepare the software engineering workforce.

## 1 INTRODUCTION

Many recent software engineering (SE) graduates often face difficulties when beginning their professional careers, e.g., according to [1] and hired.com/skills-gap. Some in the community believe that: "*The software engineering shortage is not a lack of individuals calling themselves "engineers", the shortage is one of quality – a lack of well-studied, experienced engineers with a formal and deep understanding of software engineering*" (goo.gl/MVwcqX). Many SE university programs have evolved from computer science programs and therefore still put a strong focus on theoretical and technical computer science topics as well as mathematical foundations, which seems to cause a discrepancy between the skills learned in SE university education versus those needed in the SE practice.

The authors themselves are all active SE educators and have been teaching various SE courses for more than 15 years each. Also, the authors have had active industry experience or have worked in close collaborations with practitioners in joint industry-academia projects. According to feedback from our industry partners who have hired our students, feedback from their recently-graduated students and also needs of our university departments and SE programs, we decided to conduct a systematic literature review (SLR) to synthesize the findings of various studies which have been conducted to align SE education with industrial needs.

We used the established process for performing systematic literature review (SLR) studies in SE [2] and systematically gathered a set of 33 papers on this subject published between 1995-2018. Our review aims at identifying the most important skills in industry, and synthesizing evidences of knowledge deficiencies in graduating SE students.

Our goal in this paper is to shed light on importance and knowledge gaps of different SE topics, and to ultimately answer the question of how to best train software engineers of tomorrow. By summarizing what we as a community know in this area, our article aims to benefit the readers (both educators and hiring managers) by providing a "big picture" on state of the community w.r.t. aligning SE education with industrial needs, and an "index" to the body of knowledge in this area. Let us start with a brief review of our review procedure and then continue with the findings.

## 2 THE REVIEW PROCEDURE

In our review and mapping, we followed the established process for performing systematic literature review (SLR) studies in SE [2] and also used our experience in conducting SLRs in the past [3]. We performed the searches in the Google Scholar database. All the authors conducted all the steps as a team. Our search string was: *(educational needs OR knowledge needs OR desired skills OR essential competencies OR knowledge requirements OR skill requirements) AND (software engineers OR software developers)*.

The review aimed at addressing the following Review Questions (RQs):

- **RQ 1-** What skills are the most important in the software industry? Also, given the fast-changing nature of SE, we wanted to know if the most important have changed in the last five years.
- **RQ 2-** Is there evidence of knowledge deficiencies in graduating SE students? And what are the topics with highest knowledge deficiencies?
- **RQ 3-** To what extent are soft skills important, in addition to hard (technical) skills?

We only included the papers which focused on "aligning" SE education with industrial needs, and which were based on empirical data, such as survey results or interview data. We included the latter criteria to exclude papers which were based on pure personal opinions. After compiling an initial pool of 94 papers, a systematic voting was conducted using the above criteria among the authors, and our final pool included 33 papers.

Our research method is meta-analysis [4] which is a form of synthesis that combines the quantitative data from primary studies (the pool of 33 papers in this work). The purpose of meta-analysis is to aggregate the results of primary studies to provide a consolidated big picture on a given topic.

A more detailed description of our SLR process is provided in the Web Extras of the article (goo.gl/TBEQwu). In addition, we discuss in the Web Extras how we identified and addressed the potential threats to validity to our review. In Table 1, we show the 33 papers in our final pool. All the data that we have extracted from the papers can be found in an online repository in the form of as a Google spreadsheet (goo.gl/kpx6c7). In this article, we use the "[Pi]" format to refer to the papers in the pool as listed in the online repository, as shown in Table 1. As we can see, the attention level on this topic has risen in recent years.

**Table 1-List of the studies reviewed in this meta-analysis**

| ID | Paper title |
|---|---|
| [P1] | T. C. Lethbridge, "A survey of the relevance of computer science and software engineering education," in Conference on Software Engineering Education, 1998, pp. 56-66. |
| [P2] | I. C. Mow, H. Sasa, F. Maua-Faamau, E. Mauai, M. Tanielu, "An Evaluation of Relevance of Computing curricula to Industry Needs", in Systemics, Cybernetics and Informatics, vol. 13, no. 1, pp. 7-12, 2015. |
| [P3] | B. Kitchenham, D. Budgen, P. Brereton, P. Woodall, "An investigation of software engineering curricula," Journal of Systems and Software, vol. 74, no. 3, pp. 325-335, 2005. |
| [P4] | A. Deak, G. Sindre, "Analyzing the importance of teaching about testing from alumni survey data," Norwegian informatics conference (NIK), 2013. |
| [P5] | C. Watson, K. Blincoe, "Attitudes towards software engineering education in the New Zealand industry," in Annual Conference of the Australasian Association for Engineering Education, 2017, pp. 785-792. |
| [P6] | R. Colomo-Palacios, C. Casado-Lumbreras, P. Soto-Acosta, F. J. García-Peñalvo, E. Tovar-Caro, "Competence gaps in software personnel: A multi-organizational study," Computers in Human Behavior, vol. 29, no. 2, pp. 456-461, 2013. |
| [P7] | M. E. McMurtrey, J. P. Downey, S. M. Zeltmann, W. H. Friedman, "Critical skill sets of entry-level IT professionals: An empirical examination of perceptions from field personnel," Journal of Information Technology Education: Research, vol. 7, pp. 101-120, 2008. |
| [P8] | D. M. Lee, E. M. Trauth, D. Farwell, "Critical skills and knowledge requirements of IS professionals: a joint academic/industry investigation," MIS quarterly, pp. 313-340, 1995. |
| [P9] | K. Jones, L. N. Leonard, G. Lang, "Desired Skills for Entry Level IS Positions: Identification and Assessment," Journal of Computer Information Systems, vol. 58, no. 3, pp. 214-220, 2018. |
| [P10] | R. D. Howard, "Does the information systems curriculum meet business needs: Case study of a southeastern college," PhD thesis, Capella University, 2017. |
| [P11] | C. Scaffidi, "Employers' need for computer science, information technology and software engineering skills among new graduates," International Journal of Computer Science, Engineering and Information Technology, vol. 8, no. 1, pp. 1-12, 2018. |
| [P12] | F. Patacsil, C. L. S. Tablatin, "Exploring the importance of soft and hard skills as perceived by IT internship students and industry: A gap analysis," Journal of Technology and Science Education, vol. 7, no. 3, pp. 347-368, 2017. |
| [P13] | A. Radermacher, "Evaluating the gap between the skills and abilities of senior undergraduate computer science students and the expectations of industry," PhD thesis, North Dakota State University, 2012. |

## 3 DATA FROM 12 COUNTRIES AND OVER 4,000 RESPONDENTS

Most of the papers in the pool had extracted data from one country only, e.g., [P2] had data from the UK, and [P8 … S11] had data from the USA. The advantage of our meta-synthesis (meta-analysis) is that the combined dataset has data from 12 countries which provides a stronger evidence on the subject, than each of the single studies. The top countries from which data were gathered were: USA (15 papers), Canada (4), South Africa (4), New Zealand (2) and Spain (2). Each of the following countries was represented in one paper each: UK, Norway, Philippines, Jordan, Australia, Finland and Samoa. Two papers had data from both USA and Canada and one paper surveyed data from world-wide scale.

Also, in terms of number of data points (respondents of surveys), studies had between 8 [P28] and 600 respondents [P21]. Since the studies were mostly conducted in different countries, there is slim chance that a single software engineer had participated in more than one study in the pool. Thus, when we add up the number of respondents from all 33 studies, we can say that the data and evidence are from "up to" 4,132 respondents.

By combining data and evidence from all previous studies and by having such a large combined dataset, our study aims to provide a more comprehensive view on the topic.

## 4 THE MOST IMPORTANT SKILLS IN THE INDUSTRY (RQ 1)

The questionnaires designed and used by the studies had differences w.r.t. the concrete SE topics used in them. In other words, when asking respondents to rate (rank) the importance of SE topics, different papers used different sets of SE topics. Six studies used the SE topics as proposed in different versions of the Software Engineering Body of Knowledge (SWEBOK) [5] (v1.0 developed in 1999, v2 in 2004 and v3 in 2014) [P1, S3, S4, S6, S14, S32]. Two studies [P3, S26] used a similar guideline by the IEEE, named the Software Engineering Education Knowledge (SEEK), developed in 2004. [P4] used the ACM SE 2004 curriculum guideline. [P26] used the ACM Body of Knowledge of Computing Curriculum for Computer Science (CCCS). Three other studies used the ACM IT curriculum and three used the ACM IS curriculum. The remaining 20 studies did not use a single curriculum model, but instead synthesized the list of SE topics either from the literature or by an initial interview with practitioners.

With such a diversity in the list of SE topics used in the studies, we selected the most relevant model, the latest version of SWEBOK, version 3.0 and mapped the SE topics discussed in the papers to the 15 "knowledge areas" (KAs) of SWEBOK which are: requirements, design (and architecture), development (programming), testing, maintenance, configuration management, project management, SE process, SE models and methods, quality, SE professional practice, SE economics, computing foundations, engineering foundations, and mathematical foundations.

The next step was to consolidate the quantitative data of skill (topic) importance from all the papers. Almost all papers had presented ranking of the most important skills. To be able to cross-compare and synthesize data in a consolidated manner, we harmonized the importance ranking data as follows. We normalized the ranking of topics in each paper to the range of [0, 1], for each SWEBOK KA. For example, for [P1], three of the 14 ranked topics were related to the "design" KA: general architecture (in rank 1), object-oriented design (rank 9), and user-interface design (rank 12). We calculated the average of (1, 9, 12), which equals 7.33, and divided it by the number of all SE topics in that paper (14), and the normalized rank metric was 0.52. Note that since rank data were used, the lower the value of this metric, the higher the importance of a given topic. Thus, by calculating "1- normalized rank", we aggregated the normalized importance (0.48 for the above example). Once we had the importance metric of each KA for each paper, we calculated its average among all papers.

Figure 1 show the most important topics as a scatter plot, in which the normalized importance metrics of each topic and the number of papers that it has appeared in, are shown. Also, given the fast-changing nature of the field of SE and the knowledge areas in our field, we were curious to compare the skill importance data from all the papers, versus the papers published in the last five years. Thus, we calculated the above metrics for each case separately.

Comparison of the two charts in Figure 1 provides interesting insights. When looking at all the papers, requirements, design and testing are the most important and also most frequently-mentioned topics with SE professional practice, project management and development coming next. However, when looking at the recent papers, the top-3 topics change to: SE professional practice, project management, and testing. This seems to denote that less technical skills such as SE professional practice and project management have become even more important in SE education in recent years. SE professional practice covers topics such as professionalism, group dynamics and communication skills and these are (soft) skill especially required in modern agile software development that is more strongly based on communication and interaction than traditional waterfall approaches. According to our experience, an effective approach to cover project management and SE professional practice in our education practice is by larger SE projects done by student teams either in class or even together with companies [6].

Mathematical and engineering foundations as well as SE economics are ranked low in both charts. This may highlight the establishment of SE (and its education) as a separate engineering discipline that relies on other sciences such as computer science, mathematics and economics and adopting ideas from those subjects by developing new approaches to solve problems in engineering software [7]. In line with this finding, it is also interesting to observe that requirements, testing and design are considered more important than the actual development. However, according to our experience, this is not always reflected in SE education, especially if it is embedded into computer science curricula.

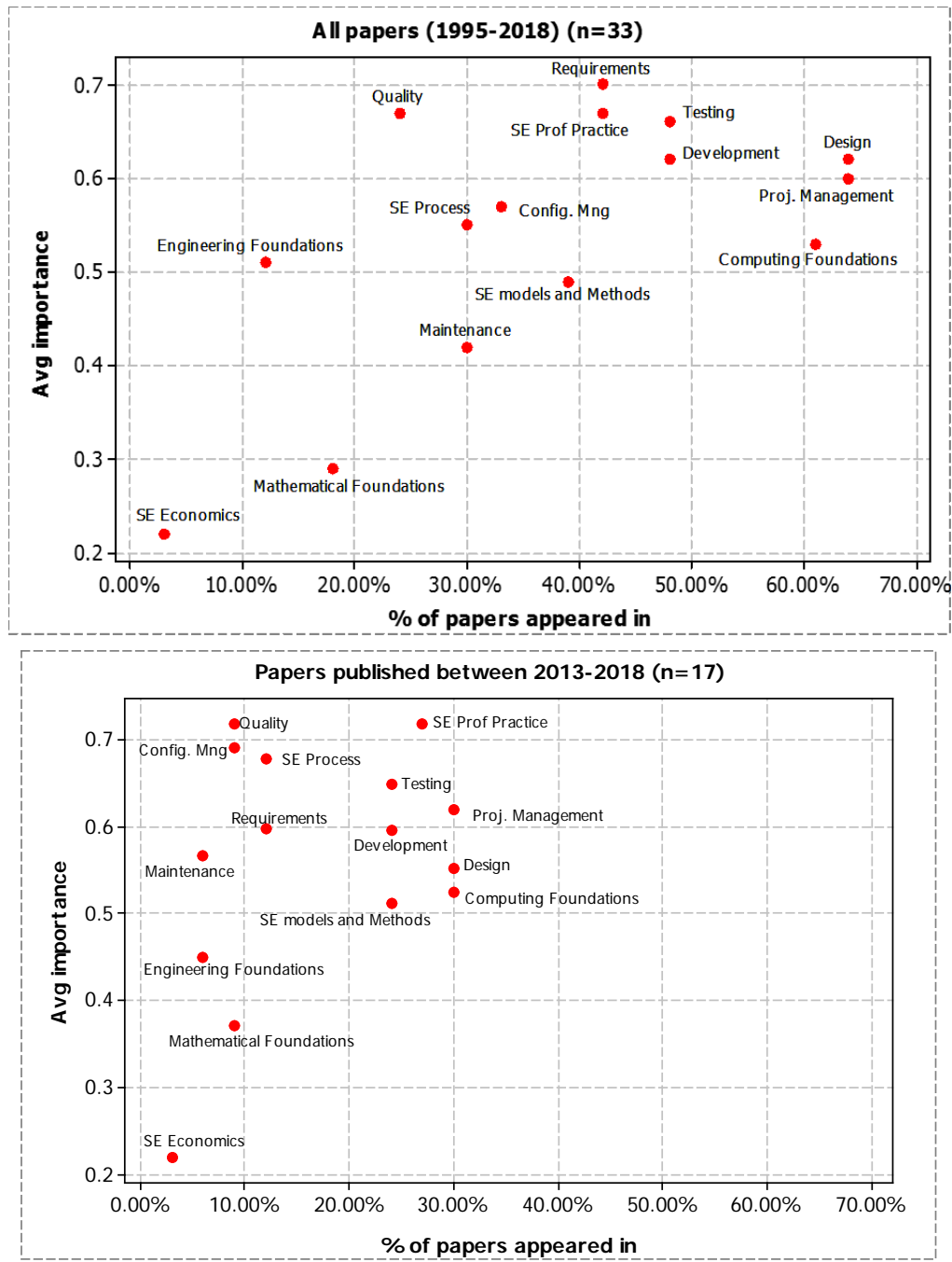

**Figure 1- The most important skills (data from all the papers, versus papers in the last five years)**

## 5 KNOWLEDGE GAPS: HIGHLIGHTING THE TOPICS THAT WE SHOULD TEACH MORE (RQ 2)

Eight (8) of the 33 studies also measured, in quantitative terms, the knowledge gap (deficiency) from their survey participants, which was usually done by subtracting the importance-in-job measure of a given SE topic from the measure of how much the participant had learnt during her/his university education. We extracted all that quantitative knowledge gap values and calculated their normalized average. We show in Figure 2 a scatter plot to visualize the average knowledge gap values versus their importance. The X-axis shows the average importance and the Y-axis the average knowledge gap. As we can see in the case of all eight papers, generally speaking, the two factors are quite correlated and with increasing reported importance, more knowledge gap has also been reported. The greatest reported knowledge gaps are in the following areas: (1) configuration management, (2) SE models and methods, (3) SE process, (4) design (and architecture), and (5) testing. Thus, in general, more education and training focus shall be given to these topics, by university programs and also when training newly-hired staff in the industry.

We have also divided the scatter plot of Figure 2 into four quadrants to be able to clearly see the SE topics with low/high importance and low/high knowledge gap.

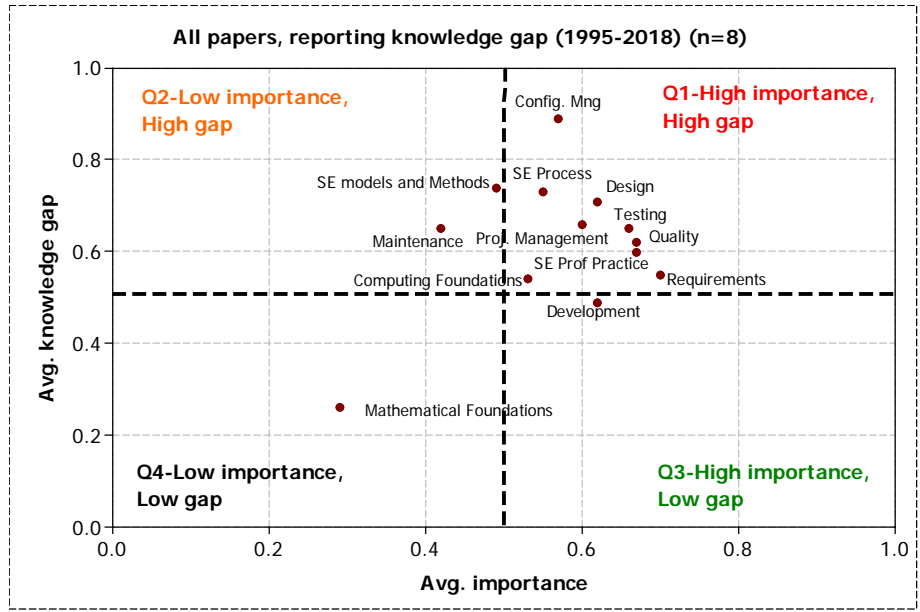

**Figure 2- Topics with the greatest knowledge gap— where importance exceeds current knowledge of survey participants**

Topics in Q1 (high importance, high gap) are those which need the highest attention w.r.t. need for improvements in SE education in university programs. They have high importance, but also high knowledge gap. Topics in Q2 (low importance, high gap) should get the next level of attention w.r.t. SE education (after those in Q1). They have relatively low importance, but there are still high knowledge gaps in those topics, and thus need for more education and training on those topics.

For topics in Q3 (high importance, low gap), the university programs are generally doing a good job, since knowledge gaps in those topics are relatively low, while they are quite important w.r.t technical needs in the industry. Only the software "development" topic slightly falls in Q3 in one of the scatter plots in Figure 2.

Topics in Q4 have low importance, and low knowledge gap, thus they are the least critical w.r.t. need for improvements and attention of SE education in university programs. The KA "Mathematical foundations" falls into Q4 in both scatter plots.

## 6 HARD SKILLS ALONE ARE NOT ENOUGH: HAVE YOU GOT SOFT SKILLS? (RQ 3)

It is widely discussed in the community that hard (technical) skills alone do not make a great software engineer [8] and soft skills are equally important (if not more). Hard skills are comprised of domain knowledge and technical skills, while soft skills are comprised of team and interpersonal skills. "*Soft skills contribute significantly to individual learning, team performance, client relations and awareness of the business context*" [P16].

24 of the 33 studies recognized the importance of soft skills. We categorized soft skills as follows: Teamwork and communication (discussed in 19 studies), Leadership (13 studies), Critical thinking (11 studies) and Others (17 studies). Many other important soft skills were also mentioned in the papers, such as: cultural fit, understand business drives, aptitude, attitude, coping with ambiguity, learning and curiosity, passion / drive to innovate, etc.

One of the studies, [P16], specifically focused on industry expectations of soft skills in IT graduates. The data came from a regional survey in New Zealand in 2016. Key findings the study of interest to educators were: While in-house technical training is widely used to advance graduate skills and teach new technologies, most employers consider these soft skills to be untrainable in the work-place, making soft skills the critical hurdle for employment. Furthermore, it was found that short-term pressure on employers for technical skills can result in the need for soft skills being overlooked. Two of the interesting quotes in the study were: "*The public sector especially needs engineers with a sophisticated understanding of the social environment within which their activity takes place, a systems understanding, and an ability to communicate with stakeholders*"; and that: "*Today's working environment is all about relationships, both internal and external. We need people who can step up and be accountable without always needing a coach/mentor standing by. People working in isolation contribute more errors than teams*".

Some studies even reported quite bold findings, e.g., survey data of an American study [P9] showed that "*soft skills are significantly more important than hard skills for entry-level positions*". A study performed in New Zealand [P5] reported that: "*Soft skills are critical skills in SE and make up seven of the top eight most important skills [in that study]*". [P22] also recommended that: "*Soft skills and business skills must be included in curricula*".

These statements are in line with the finding that SE professional practice is of high importance (see Figure 1) which comprises of topics such as professionalism, group dynamics and communication skills.

## 7 OTHER INTERESTING FINDINGS

We observed a lot of other interesting findings when reviewing the papers. For example, there were suggestions on decreasing emphasis on certain topics in SE university education (i.e., what we should teach less). [P1] expressed that "*Participants felt that their university education gave them a much better grounding in mathematics than in software topics*", and thus recommended that: "*emphasis on certain mathematics topics should be changed [decreased]*". The empirical data also showed that "*much mathematics is being forgotten, whereas much new software knowledge is being acquired on-the-job*". [P3] also reported that there is "*over-emphasis on mathematical topics and under-emphasis on business topics*" in SE education. [P3] called for less educational focus on parsing and compiler design, formal specification methods, digital electronics and digital logic in SE programs.

Going further, some studies discussed that determining how much coverage each SE topic should have is not enough, but educators should teach using "real-world" example software systems. For example, [P27] reported that "*Real-life and practical experience must be included in students' education*". [P26] also highlighted the need for "*more exposure to real life, exercises, team assignments or industry projects*". Some of the authors have had experience in such ideas [6].

Other interesting suggestions were made in [P28]: "*Instead of a greenfield project, a more valuable experience would provide students a large pre-existing codebase to which they must fix bugs (injected or real) and write additional features. Also valuable would be a management component, where students must interact with more experienced colleagues (students who have taken the class previously, who can act as mentors) or project managers (teaching assistants) who teach them about the codebase, challenge them to solve bugs several times until the "right" fix is found, or who give them sometimes capricious and cryptic weekly commandments on requirements or testing that they must puzzle out and solve together as a team*". The authors of this article often heard similar comments when talking to experienced SE practitioners.

## 8 IMPLICATIONS AND ROAD AHEAD

The findings presented in this article show the importance of SE professional practice and soft skills in general, the importance of certain SE activities and skills in SE education (especially requirements engineering, design and testing), knowledge gaps in specific areas of SE (especially configuration management, SE models and methods as well as SE process), and the importance of real-world examples in SE courses.

The authors have already started to benefit from the findings of the presented review and meta-analysis study in their SE education activities in several ways. This review has helped us to identify the most important SE topics, based on the largest synthesized body of evidence in the literature. Also, we found that the greatest knowledge gaps are in configuration management, SE models and methods, SE process, design (and architecture), and testing. Furthermore, in our ongoing university SE courses, we have started to align our teaching materials with the important topics and areas which have the greatest knowledge gaps. Also in the context of a large software company in Turkey, with which one of the authors was affiliated with, an industrial training program for potential new hires was recently conducted [9] based on the insights provided by this review study. We are certain that the results and findings presented in this paper will also benefit other educators and hiring managers by helping them adapt their education / hiring efforts to best prepare the SE workforce.

Finally, the findings also show that mathematical and engineering foundations are often overemphasized in SE programs. This highlights the need to further establish SE as a separate engineering discipline using knowledge from computer science and other basic sciences such as mathematics, economics or even psychology, and to further separate computer science from SE university programs [10].

## AUTHOR BIOGRAPHIES

Vahid Garousi is an associate professor at Wageningen University. Previously, he was an Associate Professor of Software Engineering in Hacettepe University in Ankara, Turkey (2015-2017) and an Associate Professor of Software Engineering in the University of Calgary, Canada (2006-2014). His research interests include software testing, empirical software engineering, and education research. He received a PhD in Software Engineering from Carleton University. Vahid was an IEEE Computer Society Distinguished Visitor from 2012 to 2015. Contact him at vahid.garousi@wur.nl

Görkem Giray is a software engineer and an independent researcher. His research interests include software engineering, and education research. He received a PhD in Computer Engineering from Ege University. Contact him at gorkemgiray@gmail.com

Eray Tüzün is a faculty member at Bilkent University. Before moving to academia, he worked for more than 15 years in software industry. His research interests include software analytics, software reuse, empirical software engineering, and software engineering education. He received a PhD in Information Systems from Middle East Technical University. Contact him at eraytuzun@cs.bilkent.edu.tr

Cagatay Catal is a faculty member at Wageningen University. His research interests include software engineering, education research, and data mining. He received a PhD in Computer Engineering from Yıldız Technical University. Contact him at cagatay.catal@wur.nl

Michael Felderer is a professor at the University of Innsbruck's Institute of Computer Science and a senior researcher at the Blekinge Institute of Technology. His research interests include software quality, processes and analytics as well as empirical methods and education research in software engineering. He works in close collaboration with industry and received a PhD in Computer Science from the University of Innsbruck. Contact him at michael.felderer@uibk.ac.at